\documentclass[]{aa}
\usepackage{graphicx}

\newcommand{\be}{\begin{equation}}
\newcommand{\ee}{\end{equation}}
\newcommand{\bea}{\begin{eqnarray}}
\newcommand{\eea}{\end{eqnarray}}
\begin{document}
   \title{Solar wind induced magnetic field around the unmagnetized Earth}

\subtitle{}
    \author{G.T. Birk$^{1}$, H. Lesch$^{1}$, \and C. Konz$^2$}
   \institute{$^1$ Institut f\"ur Astronomie and Astrophysik, 
 Universit\"at M\"unchen, Scheinerstr 1, D-81679 M\"unchen, Germany\\
$^2$Max-Planck-Institute for Plasma Physics, Garching, Germany }
              
 \date{}
\mail{birk@usm.uni-muenchen.de}

\abstract{The Earth is a planet with a dipolar 
magnetic field which is 
agitated by a magnetized plasma wind 
streaming from the Sun. The magnetic field 
shields the Earth's surface
from penetrating high energy solar wind particles, as well as 
interstellar cosmic rays. The magnetic dipole has reversed sign 
some hundreds of times over the last 400 million 
years. These  polarity reversals correspond to drastic breakdowns
of the dynamo action. The question arises what the consequences
for the Earth's atmosphere, climate, and, in particular, biosphere are.  
It is shown by kinematic estimates and three-dimensional
plasma-neutral gas simulations
that the solar wind can induce very fast a magnetic field in the
previously completely unmagnetized Earth's
ionosphere that is strong enough to protect Earth from cosmic
radiations comparable to the case of an intact magnetic dynamo.
}
\authorrunning{Birk, Lesch and Konz}

\titlerunning{Magnetic Field around the Unmagnetized Earth}
 \maketitle
\section{Introduction}

Paleomagnetic records show that the magnetism of Earth has
reversed itself hundreds of times over the last 400 million 
years (Valet \& Meynardier 1993; Juarez et al. 1998; Gee et al. 2000; Selkin
\& Tauxe 2000; Valet 2003).
In fact, geomagnetic polarity
reversals represent the most dynamic feature of the Earth's magnetic
field.  The polarity reversals do not occur instantaneously. Rather, 
transition periods, that span some thousand years and are characterized
by unstable varying magnetic fields with no clear shape, lay between the
stable dipole field states. 
During the transition periods the magnetic field strength can drop
well below 10$\%$  of the average value (Juarez et al. 1998; Guyodo \&
  Valet 1999; Selkin \& Tauxe 2000)
which signifies a drastic breakdown of the Earth's dynamo.
In the present time the magnetic south pole has wandered over more than
1100 km during the last 200 years. The strength of the Earth's field
has decreased by $5\%$ per century. This decrease is by far the
fastest that has been verified since the last total field reversal, the
end of the so-called Matuyama chron, 730,000 years ago. 
Also, by statistical estimates the Earth's dynamo is overdue for a reversal.
Thus, we have to expect a transition period characterized by a very
small Earth's magnetic field in the near future. Since the Earth's
dipole field provides us with a shield against cosmic rays and solar
high-energy radiation one may wonder about the consequences for life
on Earth. 
Also, having in mind that Mars lost the atmosphere almost completely
after the final breakdown of the magnetic field \cite{luh2}, 
one may speculate that 
stripping by the solar wind could alter the Earth's atmosphere.
Severe climate changes could result.
Interesting enough, during the last Brunhes-Matuyama reversal,
no major changes in plant and animal life have been 
detected. This may be partly due to the fact that the atmospheric
layers block some fraction of the cosmic radiation by scattering. 
On the other hand, the
cosmogenic radionuclide production varies at least over the last
200,000 years as function of short-term variations of the magnetic
field \cite{frank00}. 

In this contribution. we consider the interaction of the solar wind
with a completely unmagnetized Earth. 
When the solar wind encounters unmagnetized objects, such as 
Venus \cite{luh95} and comets \cite{konz}, magnetic barriers  
and ionopauses develop. Although the  
interaction 
of a fully ionized and a weakly ionized 
gas is very complex, an important 
characteristic can be identified - the generation of 
magnetic fields caused by relative 
plasma-
neutral gas shear flows. It has been 
shown \cite{hub93} that this process operates in the 
Venus' ionosphere and is responsible for 
the non-dipole magnetic field measured there.
The same process has been studied in detail for
the generation of seed magnetic fields in emerging
galaxies (Wiechen et al. 1998; Birk et al. 2002) and in circumstellar disks \cite{birk03}.
A kinematic estimate
indicates that relatively strong magnetic fields are generated in 
the Earth's ionosphere. So far, the only way to study the dynamical
non-linear interaction of the magnetized fully ionized solar wind
plasma and the partially ionized Earth's ionosphere are
three-dimensional plasma-neutral gas simulations.
Our numerical studies show the draping of the magnetic field of the solar wind
and the self-generation of a relatively strong
magnetic field in the Earth's ionosphere.

\section{A kinematic estimate of the ionospheric magnetic field
induced by the solar wind}

The interaction of a fully ionized plasma with a partially ionized gas
can be described by the fluid balance equations for the mass densities,
momentum densities and thermal pressures of the different species (see 
Sec.~3) together with the generalized Ohm's law. Ohm's law connects the 
electric fields
and electric currents in a plasma. For the low-frequency dynamics we are 
interested in, it can be deduced from the
inertialess electron momentum equation (Mitchner \& Kruger 1973; Wiechen et
al. 1998)
\bea
0 = - \nabla p_{\rm e} - n_{\rm e} e \left({\bf{E}} + {1 \over c} {\bf{v}}_{\rm e} \times
{\bf{B}}\right) - \rho_{\rm e} \nu_{\rm ei} ({\bf{v}}_{\rm e} - {\bf{v}}_{\rm i}) -
\nonumber\\
\rho_{\rm e} \nu_{\rm en} ({\bf{v}}_{\rm e} - {\bf{v}}_{\rm n})\label{emoment}
\eea
where ${\bf E}$ and  ${\bf B}$ are the electric and magnetic fields,
respectively,
and by $p_{\rm e}$, $n_{\rm e}$, $\rho_{\rm e}$, $v_\alpha$, and $\nu_{\alpha\beta}$ 
the electron pressure, particle density,
mass density, bulk velocities of the different plasma components
(electrons, ions, and neutrals) and collision frequencies are denoted. 
The constants $e$ and $c$ are the elementary charge and speed of light.
The electron velocity ${\bf{v}}_{\rm e}$ can be written in the form
${\bf{v}}_{\rm e} = {\bf{v}} -  m_{\rm i} \ {\bf j}  /e \rho$ with the plasma mass density 
$\rho=\rho_{\rm e}+\rho_{\rm i}\approx \rho_{\rm i}$, 
the plasma bulk velocity ${\bf v}=(\rho_{\rm 
e}{\bf v_{\rm e}} +\rho_{\rm i}{\bf v_{\rm i}})/\rho\approx {\bf
v_{\rm i}}$ and the 
electric current density ${\bf j}= ne ({\bf v_{\rm i}} - {\bf v_{\rm e}})$
($n= n_{\rm e} = n_{\rm i}$ is the particle density of the quasi-neutral 
plasma). 
Consequently, the inertialess equation of motion of the electrons yields Ohm's 
law in the form
\bea
{\bf{E}} + {1 \over c} {\bf{v}} \times {\bf{B}} = {1 \over {n e}}
\nabla p_{\rm e} + {1 \over {4 \pi n e}} (\nabla \times {\bf{B}}) \times
{\bf{B}} -\nonumber\\
 {1 \over {n e}}[\rho_{\rm e} \nu_{\rm ei} ({\bf{v}}_{\rm e} - {\bf{v}}_{\rm i})
- \rho_{\rm e} \nu_{\rm en} ({\bf{v}}_{\rm e} - {\bf{v}}_{\rm n})]\label{ohm}
\eea
where Amp\'ere's law has been used to eliminate the current density. 
By Faraday's law one obtains an equation that governs the dynamical
evolution of the magnetic field
\begin{eqnarray}
{{\partial {\bf{B}}} \over {\partial t}} & = &
\nabla \times ({\bf{v}} \times {\bf{B}}) -
{c \over e} \nabla \times \left({{\nabla p_{\rm e}} \over n}\right) -
\nabla \times (\eta \nabla \times {\bf{B}})
\nonumber \\
& & -
{c \over {4 \pi e}} \nabla \times \left({{(\nabla \times {\bf{B})} \times
{\bf{B}} } \over n}\right)\nonumber\\& &
 -
{{c m_{\rm e}} \over e} \nabla \times [ \nu_{\rm en} ({\bf{v}}_{\rm e} - {\bf{v}}_{\rm n})]
\label{induct}\end{eqnarray}
with the magnetic diffusivity $\eta=c^2\nu_{\rm ei}m_{\rm e}/4\pi n e^2$.
The final term, that results from the collisional momentum transfer between
electrons and neutrals due to shear flow, is crucial for the 
generation of the magnetic fields under investigation.
The collision frequency $\nu_{\rm en}$ is
given by \cite{huba98} $\nu_{\rm en} = n_{\rm n}\sigma (k T_{\rm e} / m_{\rm e})^{1/2}$
with a scattering
cross section of $\sigma \simeq 4 \times 10^{-15} {\rm cm}^{-2}$ and $k, T_{\rm e}$
denoting the Boltzmann constant and the electron temperature,
respectively.
The main interaction between the solar 
wind and the Earth's ionosphere
is expected to happen at a ionospheric height of 
about 350 km where the ram
pressure of the solar wind with an
average velocity of $v_{\rm SW}=450$ km/s is
balanced by the ionospheric pressure. The 
collision frequency of solar
wind electrons with ionospheric neutrals 
can be estimated to be $\nu_{\rm en}\approx 200 $Hz. 
The solar wind 
consists of a fully ionized plasma
that carries along a magnetic field of 
about $B_{\rm SW}\approx 2\times 10^{-4}$G (see, e.g., \cite{priest}
for the solar wind parameters).
The shear 
length scale $L$, that defines an estimate for the nabla operator in the
final term in Eq.~3, is the least
known parameter in our analysis. 
In fact, we scale our numerical results 
to this (free) parameter. Shear lengths of $L=10$  km seem to be
reasonable. The ionospheric height gives an upper limit for $L$. 
The order of magnitude of
the generated magnetic fields in the rest frame of the neutrals
can roughly be estimated from the final term of Eq.(3) as 
\bea
&& B_{\rm gen}  \approx  \tau \nu_{\rm en} \frac{m_{\rm e} c}{e}\
\frac{v_{\rm SW}}{L}\nonumber\\ &&
= 5\times 10^{-4} {\rm G} \left[\frac{\nu_{\rm en}}{200 
\ {\rm Hz}}\right]
\left[\frac{v_{\rm SW}}{450 \ {\rm km/s}}\right]
\left[\frac{L}{10\ {\rm km}}\right]^{-1} \tau
\label{bgen}\eea
where $\tau$ is the generation time scale.
The time scale of the diffusion of the magnetic field measured by 
$\eta$ is much larger than $\tau$. 
An upper limit for $\tau$ cannot be given from a pure kinematic 
consideration. The saturation magnetic field results from self-consistent
simulations of the entire dynamics presented in Sec.~3.

For the parameters given, a field 
strength comparable to the present
dipole value is generated after only ten minutes in the 
ionosphere. 
Thus, magnetic fields can be  generated very efficiently  around the 
unmagnetized Earth. 

\section{The numerical model}

The interaction of the solar wind with the Earth's ionosphere can be
modeled by a plasma-neutral gas two fluid description.
In our simulations, the following normalized plasma -
neutral gas equations are numerically integrated
\begin{equation}
{{\partial \rho} \over {\partial t}} = - \nabla \cdot ({\bf{v}} \rho)
\label{kontipl}\end{equation}
\begin{equation}
{{\partial \rho_{\rm n}} \over {\partial t}} = 
- \nabla \cdot ({\bf{v}}_{\rm n} \rho_{\rm n})
\label{kontin}\end{equation}
\begin{equation}
{\partial \over {\partial t}} (\rho {\bf{v}}) = - \nabla \cdot (\rho
  {\bf{v}} {\bf{v}}) - \nabla p + (\nabla \times {\bf{B}}) \times
  {\bf{B}} -\rho \nu_{\rm pn} ({\bf{v}} - {\bf{v}}_{\rm n})
\label{momentpl}\end{equation}
\begin{equation}
{\partial \over {\partial t}} (\rho_{\rm n} {\bf{v}}_{\rm n}) = - \nabla \cdot 
(\rho_{\rm n}
  {\bf{v}}_{\rm n} {\bf{v}}_{\rm n}) - \nabla p_{\rm n}
 \label{momentn}          - \rho_{\rm n} \nu_{\rm np} ({\bf{v}}_{\rm n} - {\bf v})
\end{equation}
\begin{equation}
{{\partial {\bf{B}}} \over {\partial t}} = \nabla \times ({\bf{v}}
\times {\bf{B}}) - \nabla \times (\eta \nabla \times {\bf{B}}) -
\nabla \times [\hat\nu_{\rm en} ({\bf{v}}_{\rm n} - {\bf{v}})]
\label{induktion}\end{equation}
\bea
 {{\partial p} \over {\partial t}}  & = &  - {\bf{v}} \cdot \nabla p -
\gamma p \nabla \cdot {\bf{v}} +  (\gamma - 1) \Biggl(2 \eta \left
(\nabla \times
{\bf{B}} \right)^2 \nonumber\\& &
- 3 \nu_{\rm pn} 
\left(p - {\rho \over {\rho_{\rm n}}} p_{\rm n}
\right ) + \rho \nu_{\rm pn} \left (
{\bf{v}} - {\bf{v}}_{\rm n} \right )^2 \Biggr)
\label{pressurepl}\eea
\bea&&
{{\partial p_{\rm n}} \over {\partial t}}  =  - {\bf{v}}_{\rm n} \cdot \nabla 
p_{\rm n} -
\gamma_{\rm n} p_{\rm n} \nabla \cdot {\bf{v}}_{\rm n} +\nonumber\\ &&
 (\gamma_{\rm n} - 1) \left ( 3 \nu_{\rm np} \left (p_{\rm n} - {\rho_{\rm n} 
\over {\rho}} p
\right ) + \rho_{\rm n} \nu_{\rm np} \left (
{\bf{v}}_{\rm n} - {\bf{v}}\right )^2 \right )
\label{pressuren}\eea
where $p= p_{\rm e} + p_{\rm i}$ and $\gamma$ denote the
thermal plasma pressure and the ratio of specific heats 
(all quantities are now to be taken as dimensionless).
The index $n$ identifies the neutral gas quantities. The effective
plasma-neutral gas collision frequency is 
$\nu_{\rm pn}=(m_{\rm e} \nu_{\rm en}+ 
m_{\rm i} \nu_{\rm in})/(m_{\rm e}+m_{\rm i})$ where
$\nu_{\rm pn}\rho=\nu_{\rm np}\rho_{\rm n}$ holds to guarantee
momentum conservation.
In the induction equation (Eq.~9),
$\hat \nu_{\rm en}$ denotes the normalized expression for 
$\nu_{\rm en} c m_{\rm e}/e$.
The final term in the induction equation results in 
self-magnetization caused by relative sheared plasma-neutral gas
flows (Lesch et al. 1989;  Huba \& Fedder 1993;  Wiechen et al. 1998). 
Normalization of
the physical quantities is carried out by the typical plasma mass density,
the typical length scale, the plasma Alfv\'en speed and  time
scale. The thermal pressure is normalized to the magnetic pressure. The thermal
pressure is measured by a typical number density and the kinetic temperature.
The source terms in the balance
equations for the mass densities, momenta and thermal pressures are
treated in a flux-corrected manner \cite{birk96}. 
For the simulation shown, we use a numerical domain of 7 Earth radii
around the Earth in each direction (with the exception of 3 Earth radii in
the solar direction) and one million grid points.
The Earth is located at $x=y=z=0$, the solar wind enters the numerical
box from the upper $y$-boundary. At all the other boundaries outflow
conditions are chosen. The neutral gas of the ionospheric layer around
the Earth shows a ${\rm cosh}^{-2}$-decay toward the interplanetary space.

\section{Numerical Results: Draping and self-magnetization}

Fig.~1 shows an arrow plot of the solar wind velocity field at three
different times (t=30 s; upper plot, t=60 s; middle plot 
and t=600 s; lower plot). 
The wind encounters the Earth and is deflected around
the planet. 
\begin{figure}
  \centering{
   \includegraphics[width=9.cm,keepaspectratio]{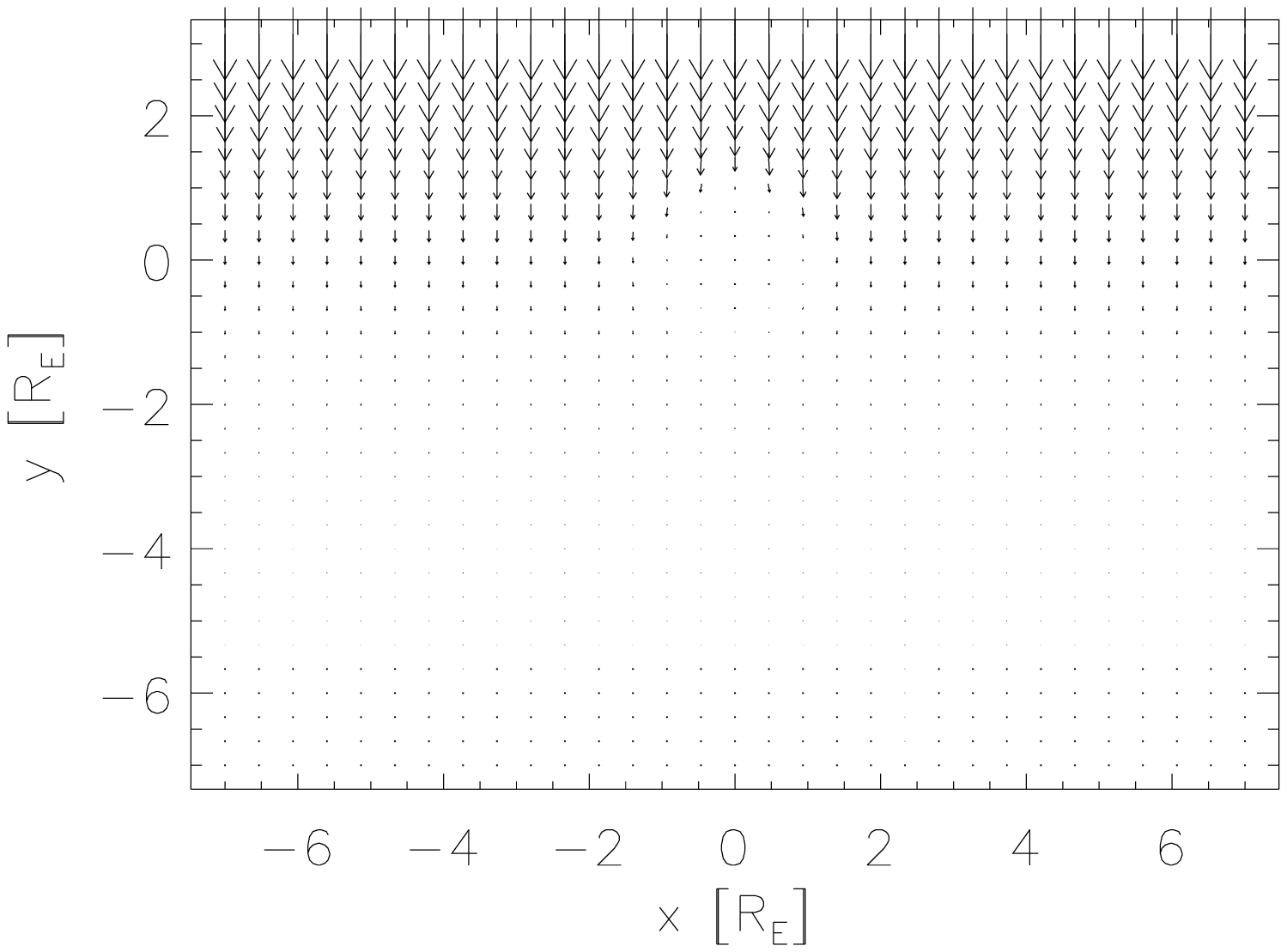}
   \includegraphics[width=9.cm,keepaspectratio]{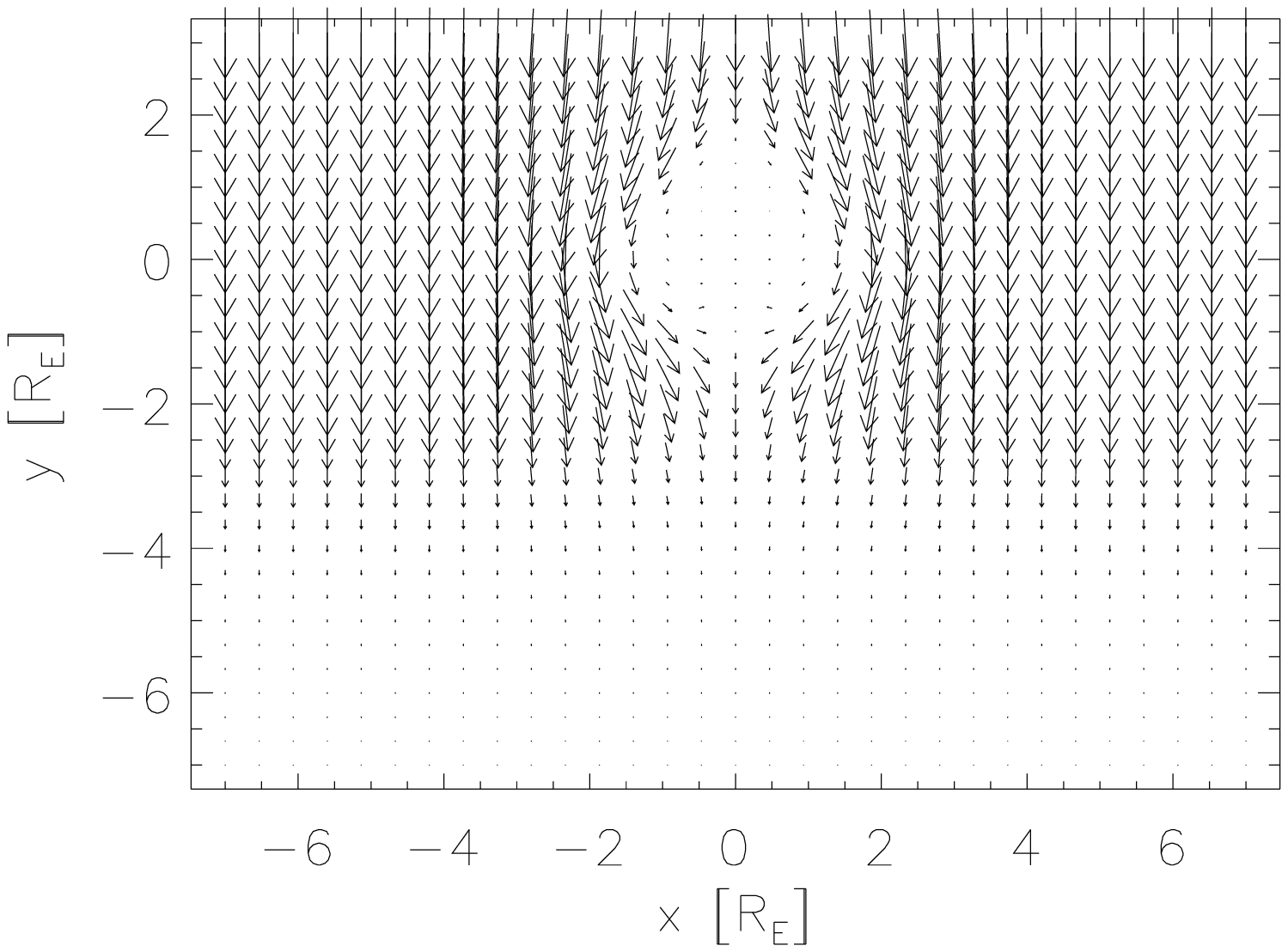}
   \includegraphics[width=9.cm,keepaspectratio]{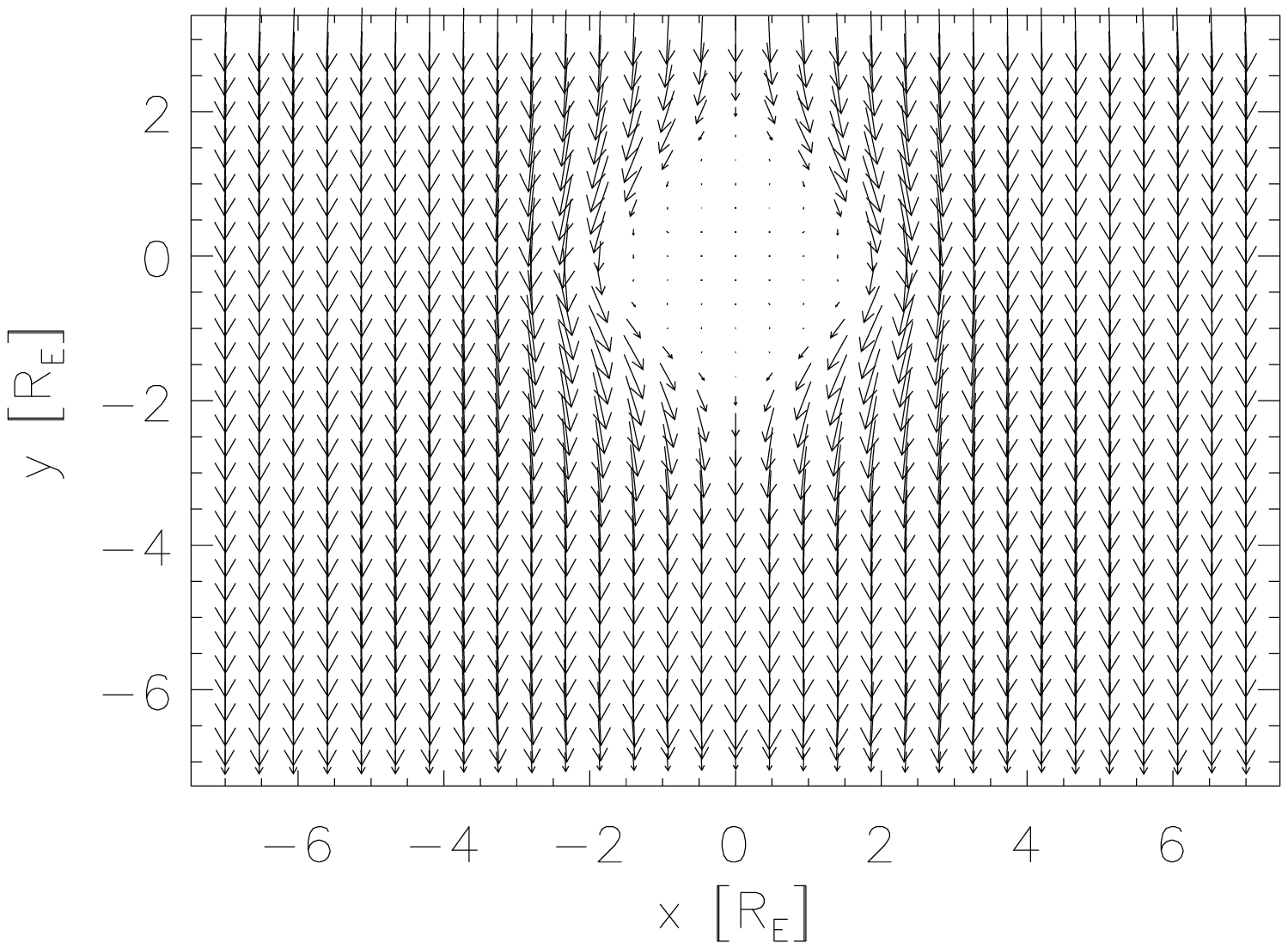}}
  \caption{The solar wind flow around the Earth in the $z=0$-plane
at t=30 s (upper plot), t=60 s (middle plot) 
and t=600 s (lower plot).}
  \label{fig1}
\end{figure}

The magnetic field lines carried by the solar wind are draped around
the Earth (Fig.~2). The draping leads to an amplification of the
magnetic field near the Earth by one order of magnitude. This effect
is well known, e.g., from investigations on the interaction of the solar wind
with the unmagnetized Venus (Russel 1993; Luhmann 1995).
 
\begin{figure}
  \centering{
\includegraphics[width=9.cm,keepaspectratio]{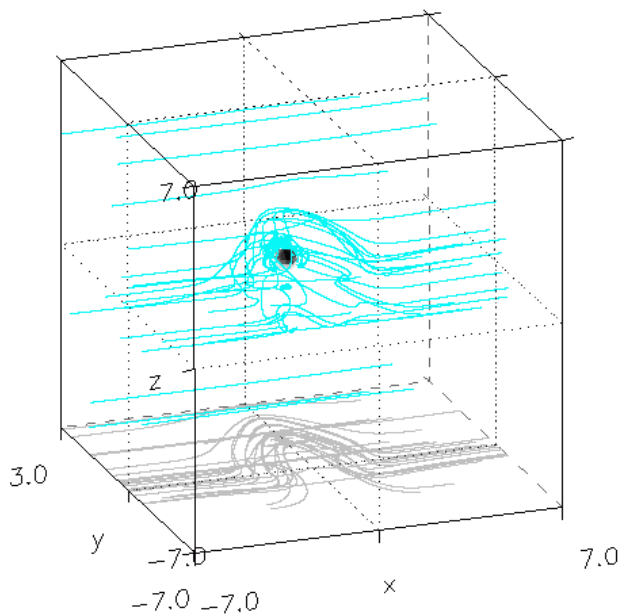}}
  \caption{The draping of the magnetic field lines of the solar wind around
    the Earth at t=600 s.}
  \label{fig2}
\end{figure}

Close to the Earth, the momentum transfer between the
charged particles of the solar wind and the neutrals of the Earth's
ionosphere becomes important 
(see final terms in Eqs.~\ref{momentpl}, \ref{momentn}).
Consequently, a new strong non-dipole 
magnetic field is generated by the sheared relative
plasma-neutral gas motion (see final term in Eq.~\ref{induktion}).
Parameter studies show the kinematic finding (see Sect.~2) that the
strength of the generated magnetic field depends on the shear length
$L$. The maximum shear length is fixed in the simulation by an appropriate
choice of the profile for $\hat \nu_{\rm en}$. 
For a shear length $L=10$ km a magnetic field of about the present
dipole strength ($B_{\rm max}\approx 0.3$ G) is induced in the 
ionosphere after about 10 minutes(Fig.~3).  
For a given $L$ the time scale of the
field generation $\tau$ results from the dynamics.
The field is generated
in heights of some hundreds of 
kilometers all around the Earth with the exception of the
subsolar region where the magnetic field is weaker.
If the shear length were chosen as say 100 km, the maximum of the generated 
field strength would be $B_{\rm max} \approx 0.03$ G.

We find that the draping effect is much weaker than the magnetic field 
self-generation by the shear flow. 

\section{Conclusions}

We studied the interaction of the magnetized fully ionized solar wind plasma 
with the unmagnetized partially ionized Earth's ionosphere. When the solar 
wind hits the Earth the magnetic field lines carried along with it are draped 
around the planet. What is more important, the relative motion between the 
solar wind plasma and the ionosphere results in the self-generation of 
magnetic fields in the ionospheric layer. The strengths of these fields 
depend on the shear length of the relative flows, which, in contrast to the 
other relevant physical parameters, is not well known. For a reasonable 
shear length of $L=10$ km the maximum strength of the newly generated 
magnetic field is comparable to the one of the present dipole field. 
Consequently, even in the case of a complete breakdown of the Earth's 
dynamo, the biosphere is still shielded against cosmic rays, in particular
coming from the sun, by the magnetic field induced by the solar wind.

\begin{figure}
  \centering{
\includegraphics[width=9cm,keepaspectratio]{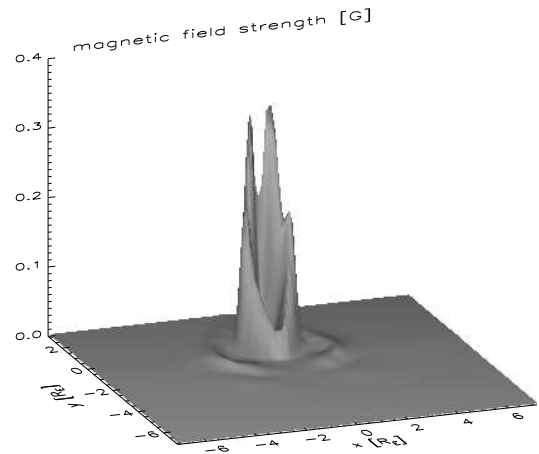}}
  \caption{The strength of the induced magnetic field in the Earth's
    ionosphere at t=500 s.}
  \label{fig3}
\end{figure}

\end{document}